\documentclass[a4paper,11pt]{article}
\usepackage{jinstpub} 
\usepackage{lineno}
\usepackage{subfig}
\usepackage{caption}

\usepackage[usenames,dvipsnames]{xcolor}
\usepackage{soul}

\proceeding{Topical Workshop on Electronics for Particle Physics 2023\\
Geremeas (CA), Italy\\
02-06 October 2023}

\sethlcolor{RedOrange}
\newcommand{\corrtex}[1]{{#1}}

\title{\boldmath Novel Developments on the OpenIPMC Project}

\author[1]{Luigi Calligaris\corrtex{,}} 
\author[1]{Carlos Ruben Dell'Aquila\corrtex{,}}  
\author[1]{Antono Vitor Grossi Bassi\corrtex{,}} 
\author[1]{André Muller Cascadan\corrtex{,}} 
\author[2]{Luis Eduardo Ardila-Perez\corrtex{,}} 
\author[2]{Marvin Fuchs\corrtex{,}} 
\author[3]{Alp Akpinar\corrtex{,}} 
\author[3]{Andrew Peck\corrtex{,}} 
\author[3]{Daniel Gastler\corrtex{,}} 
\author[4]{Giacomo Fedi} 

\affiliation[1]{Center for Scientific Computing, São Paulo State University,\\ 
Rua Dr. Bento Teobaldo Ferraz 271, Bloco II - Térreo, 01140-070 São Paulo - SP, Brazil}

\affiliation[2]{Institute for Data Processing and Electronics (IPE), Karlsruhe Institute of Technology (KIT),\\
Hermann-von-Helmholtz-Platz 1, D-76344 Eggenstein-Leopoldshafen, Germany}

\affiliation[3]{Department of Physics, Boston University,\\
590 Commonwealth Avenue, Boston, MA 02215, US}

\affiliation[4]{Blackett Laboratory, Physics Department, Imperial College London,\\
Prince Consort Rd, London, SW7 2BW, UK}

\emailAdd{luigi.calligaris@cern.ch}

\abstract{We present the recent developments in the context of the OpenIPMC project, which proposes a free and open-source Intelligent Platform Management Controller (IPMC) software and \corrtex{an} associated  controller \corrtex{mezzanine} card for use in ATCA electronic boards. We discuss our experience in the operation of OpenIPMC on prototype boards designed for the upgrades of particle physics experiments at CERN \corrtex{and we} show the addition of new features and support for new protocols in the \corrtex{firmware of the controller mezzanine card.}}


\arxivnumber{2310.19742} 

\notoc

\begin{document}
\maketitle
\flushbottom

\section{OpenIPMC and its context}
\setcounter{page}{0}
Numerous high-performance FPGA-based electronic boards have been designed for use in the upgrades of the HL-LHC\,\corrtex{\cite{ZurbanoFernandez:2020cco}} experiments for applications like detector control, calibration, trigger and data acquisition. Many of these boards are compliant with the PICMG Advanced Telecommunication Computing Architecture (ATCA) standard\,\cite{picmg_3_0_atca}. Examples of these boards are the Apollo\,\cite{Albert:2019kvt}, the Serenity\,\cite{Rose:2019oiy}, the ATLAS Trigger Processor\,\cite{Iakovidis:2023ajz} and the CMS DAQ and Timing Hub\,\cite{CMS:2022jry}. The ATCA standard requires all boards to be equipped with an Intelligent Platform Management Controller (IPMC), which monitors the health of the boards and manages their power state as part of the Hardware Platform Management (HPM) infrastructure. The OpenIPMC project\,\cite{openipmc_repo} aims to create a low-cost, free and open-source IPMC solution which is fully customizeable by the ATCA board designer. The project has been documented in previous publications and is composed of a platform-independent software component implementing the IPMI behavior (OpenIPMC-SW)\,\cite{Calligaris:2020pgl}, 
a hardware mezzanine (OpenIPMC-HW)\,\cite{Calligaris:2021ihv} suitable to be installed in the IPMC slot of a compatible ATCA carrier board and its firmware (OpenIPMC-FW)\,\cite{Calligaris:2021ihv}. \corrtex{This firmware is composed of a triplet of executable binaries designed to run on the two ARM Cortex-M7 and -M4 cores of the STM32H7 microcontroller on the mezzanine. Two binaries contain the instructions executed by the two cores in normal operation and the third one is the bootloader, which runs on the Cortex-M7 core. The firmware, which is stored into the internal memory of the microcontroller, } integrates OpenIPMC-SW with the necessary board-specific customizations and other board control services. In this document we present \corrtex{the latest developments on OpenIPMC-FW, which have mainly focused on} satisfying the requirements of the Serenity and Apollo ATCA boards \corrtex{for use in the back-end systems of LHC experiments. Both boards will be hosted in electronics shelves located in the experimental areas and operated remotely from control rooms located in the surface}.
\begin{figure}[h]
  \centering
  \includegraphics[height=25mm]{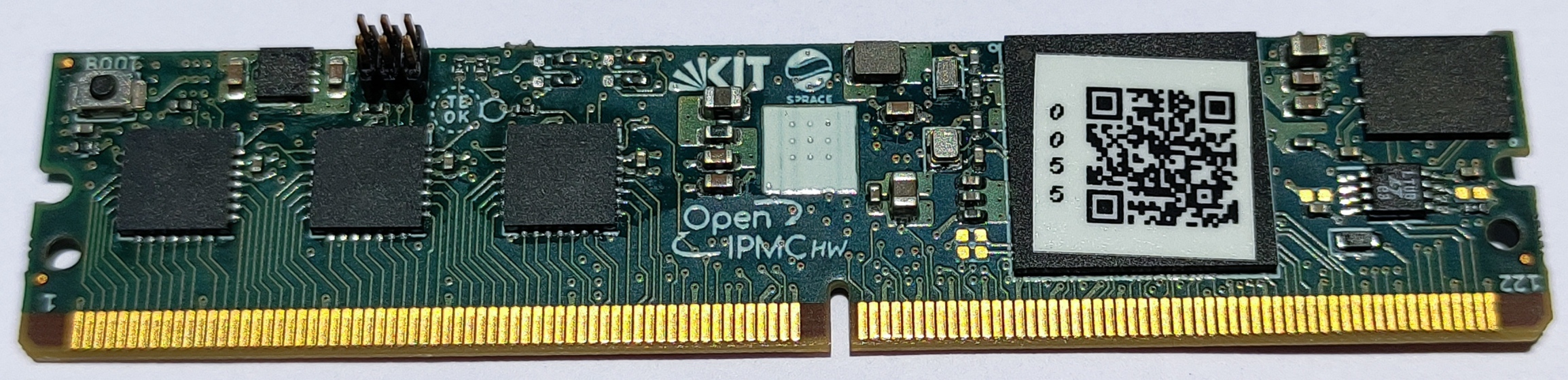}
  \caption{\corrtex{The OpenIPMC-HW mezzanine, version 1.1.}}
\end{figure}
\section{Developments on OpenIPMC-FW}
\begin{figure}[h]
  \centering
  \hfill%
  \subfloat[Breakout board being used for firmware development at SPRACE.]{\label{fig:breakout_board-4-rgb} \includegraphics[height=46mm]{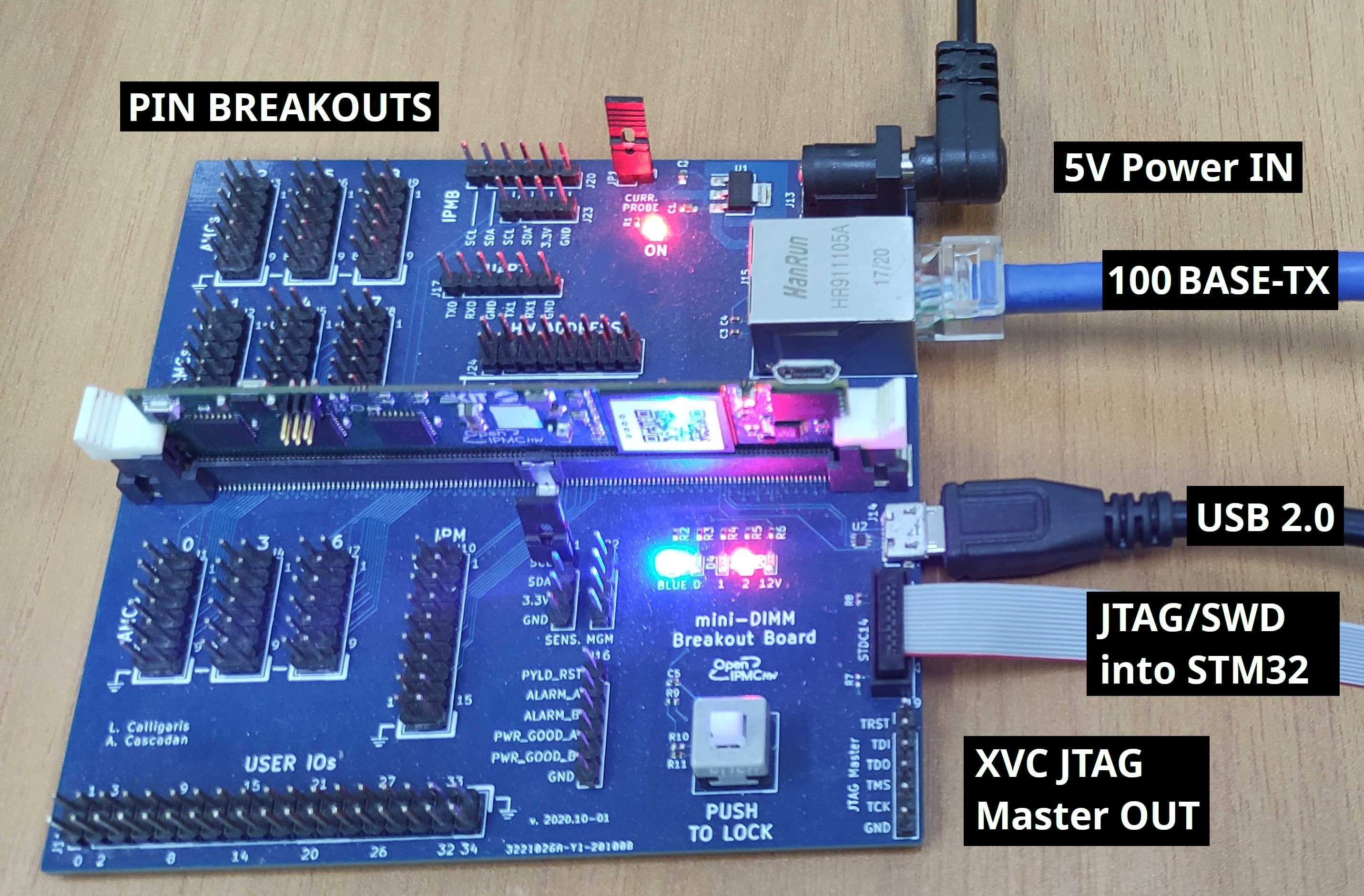}}\hfill%
  \subfloat[Backend development setup at the CERN CMS Tracker Integration Facility.]{\label{fig:atca_carriers_with_openipmc_in_b168-cut} \includegraphics[height=46mm]{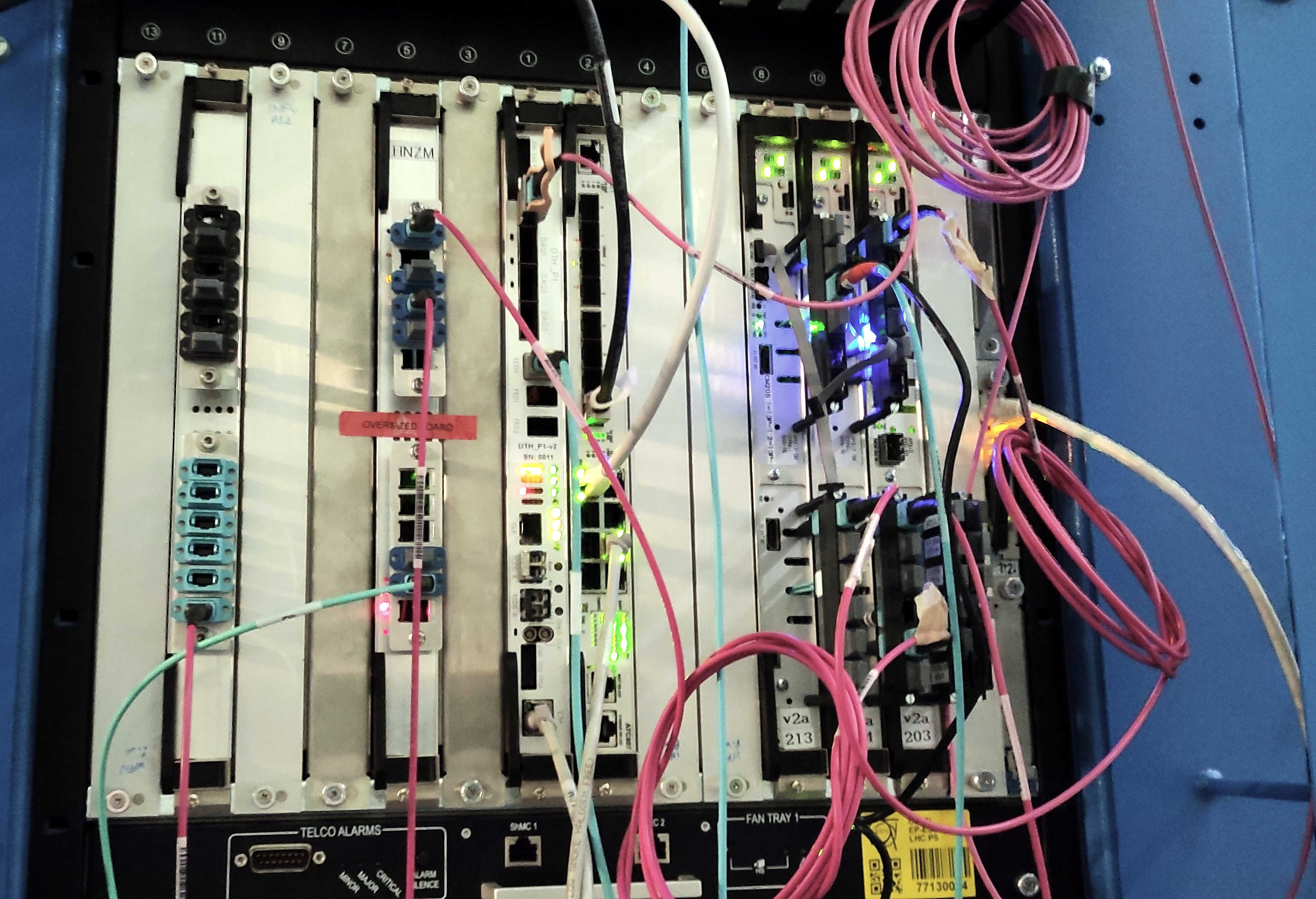}}\hfill%
  \hfill%
  \caption{Examples of OpenIPMC-HW setups used in our development.}
\end{figure}

The firmware has been subject to extensive testing in the past three years, which led to improvements in its stability and performance. Tests and development were carried out by operating the mezzanine on the bench while hosted into a custom breakout development board (Fig.\,\ref{fig:breakout_board-4-rgb}), and by operating it in Pulsar-IIb\,\corrtex{\cite{Ajuha:2017frj}}, Apollo and Serenity boards hosted in suitable ATCA shelves at the São Paulo Research and Analysis Center (SPRACE) of UNESP, at Boston University, at the Karlsruhe Institute of Technology (KIT) and at CERN (Fig.\,\ref{fig:atca_carriers_with_openipmc_in_b168-cut}). Aside from improvements to the existing firmware components, new functionalities were added and then validated for stability. \corrtex{Furthermore, some small modifications to the hardware design - which will not be described in detail here - were introduced to improve compliance with the ATCA standard, allowing to reset the IPMC without upsetting the operation of the payload when in active state.}

\subsection{Support for Xilinx Virtual Cable}
\begin{figure}[h]
  \centering
  \subfloat[\corrtex{Connection between Vivado, HW server, OpenIPMC XVC server and FPGA.}]{\label{fig:xvc-schem} \includegraphics[height=30mm]{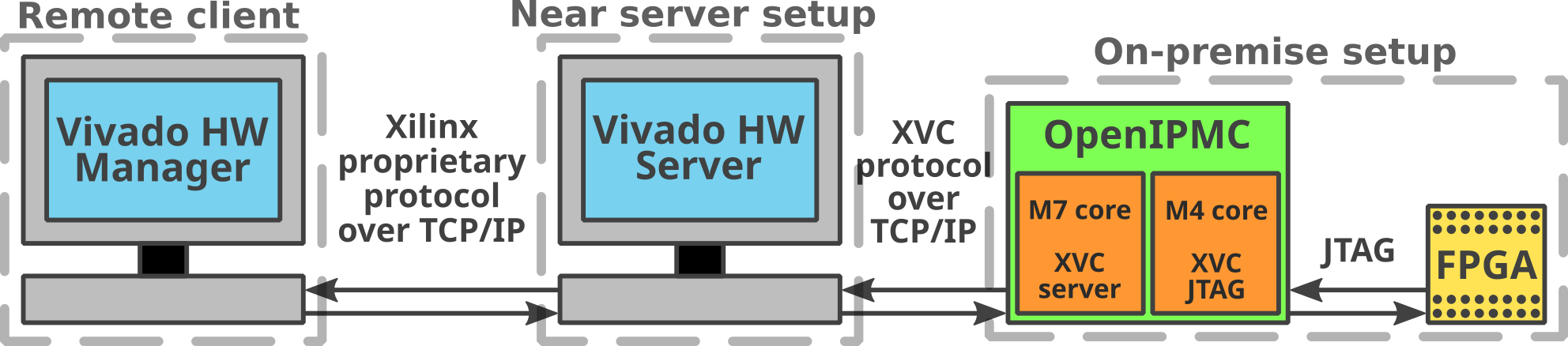}}\hfill\\
  \subfloat[XVC used to debug a Kintex-7 board.]{\label{fig:openipmc_xvc_testing-cut}\includegraphics[height=36mm]{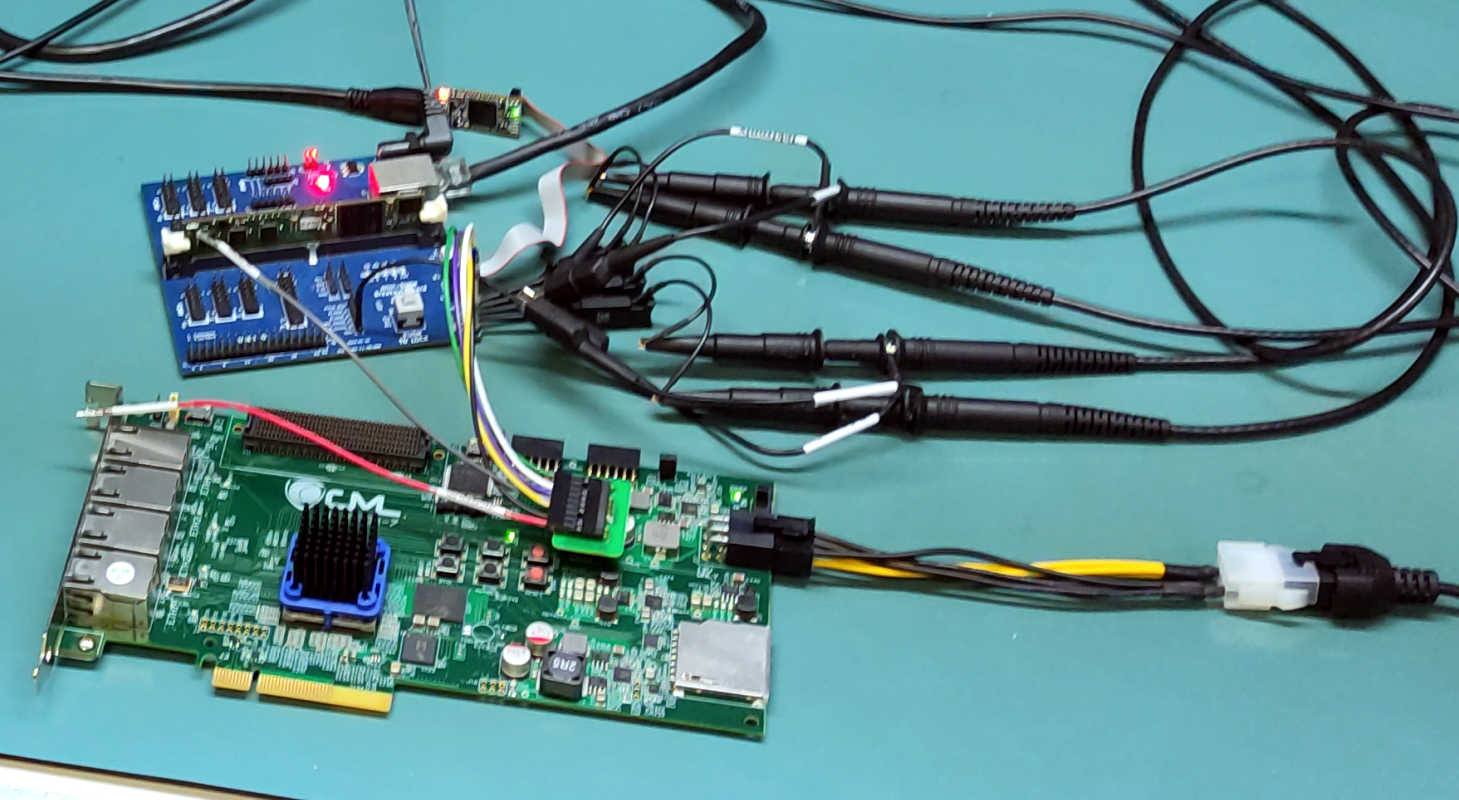}}%
  \hspace{2mm}
  \subfloat[Test pattern on TCK, TMS, TDI signals.]{\label{fig:xvc-bitbang-oscope-4-rgb} \includegraphics[height=36mm]{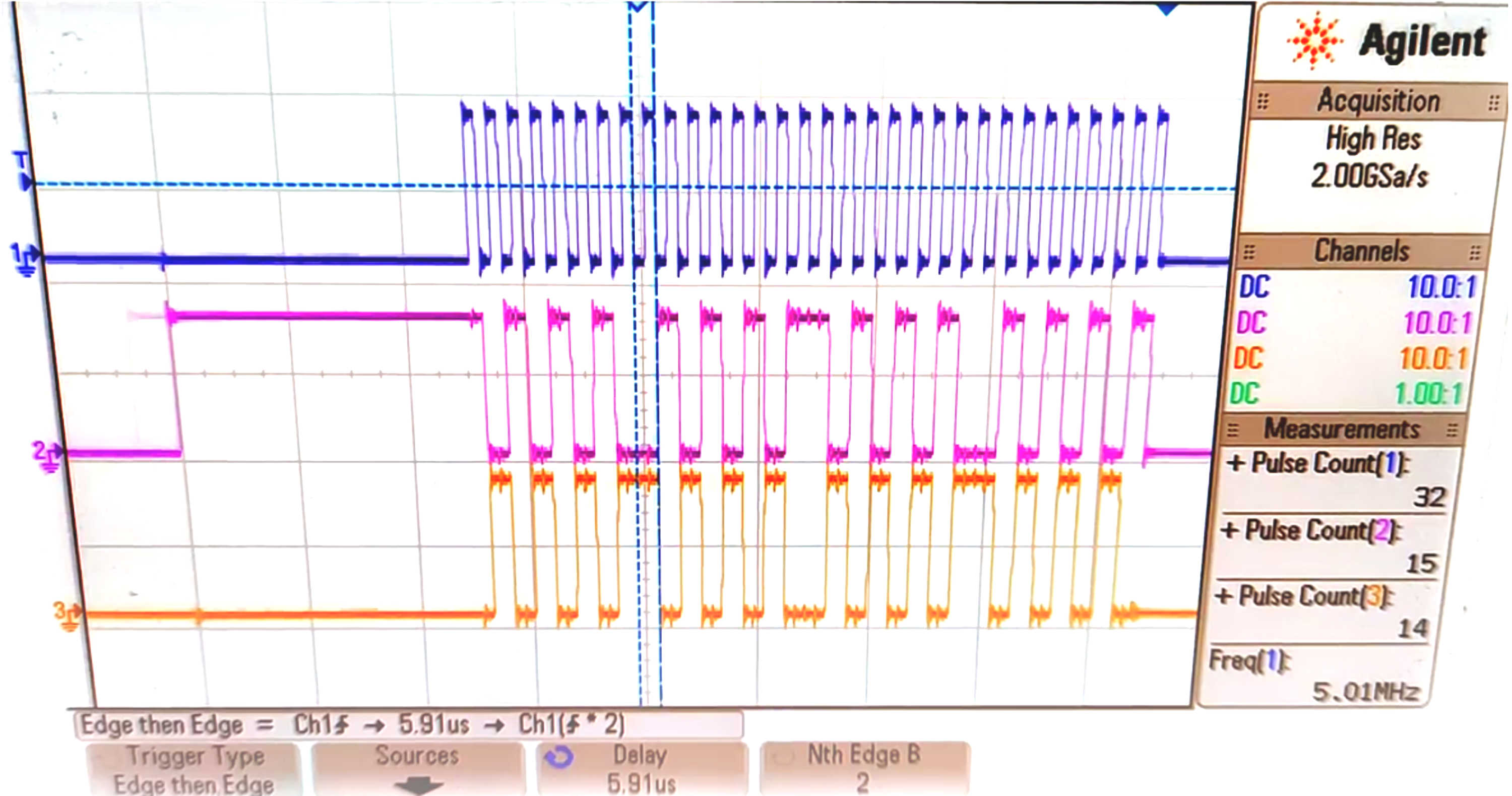}}\hfill\\
  \caption{\corrtex{Testing of the XVC JTAG server implemented in OpenIPMC.}}
\end{figure}

\corrtex{Support was added in OpenIPMC-FW for} the Xilinx Virtual Cable (XVC) protocol, which allows an instance of the Vivado development suite running on a PC to control a JTAG master interface on the IPMC over the network (Fig.\,\ref{fig:xvc-schem}). This interface, used to program and debug FPGAs and SoCs manufactured by AMD Xilinx, is broken out on a set of standardized pins on the OpenIPMC-HW edge connector, such that the board designer can route these JTAG signals to onboard devices. \corrtex{XVC support is not part of the ATCA standard but is important in the foreseen application of the IPMC in the back-end electronics of LHC experiments, as the latter make extensive use of FPGA devices.} In the OpenIPMC-FW implementation, the bit-banging and capture of the JTAG output and input signals are offloaded to a DMA controller driven by a hardware timer. This offloading solution provides very fast and consistent timing for the signals while keeping the processor resource usage low. In our tests, we were able to successfully program FPGAs of the Spartan-6 (using the ISE 14.7 suite), Artix-7 and Kintex-7 (Fig.\,\ref{fig:openipmc_xvc_testing-cut}) families, with a maximum configurable JTAG speed of 5.0 Mbps during bursts (Fig.\,\ref{fig:xvc-bitbang-oscope-4-rgb}) and an average speed of 1-2 Mbps when using XVC over a low-latency LAN connection. Once validated with these tests, the XVC server has been adopted as a development tool and is now regularly used to debug the Zynq Ultrascale+ device installed on the Apollo board. We expect that a hardware revision of the Serenity board will enable the same function in the near future.
\subsection{Support for Syslog}
A mechanism to transmit alert, debug and monitoring messages from the IPMC devices to a monitor device over the network is very useful to identify software bugs and take action in case of faults, both during the early development phase in the laboratory using a small number of devices and, later, in the production phase when operating several hundred or thousand devices. \corrtex{The Syslog protocol is not part of either the IPMI or ATCA standards but has grown into one of the most used standards in the IT industry to transport such messages over UDP/IP packets. The standard is defined in two incompatible versions by the IETF documents RFC 3164\,\cite{RFC3164} and RFC 5424\,\cite{RFC5424}. In} OpenIPMC-FW, support for both of these standards was added, with the possibility for the user to choose - at compile time - which RFC standard to use for its Syslog client. A run-time configuration parameter allows the user to choose the UDP/IP address and port of the Syslog server which will receive the messages. Following the implementation of Syslog into the firmware, the client was successfully tested to interface with an instance of Grafana Loki OSS\,\cite{GrafanaLoki}, a free and open-source log aggregator designed for cloud deployment.

\subsection{Support for RMCP/RMCP+ and Serial-Over-LAN}
\begin{figure}[h]
  \centering
  \includegraphics[height=45mm]{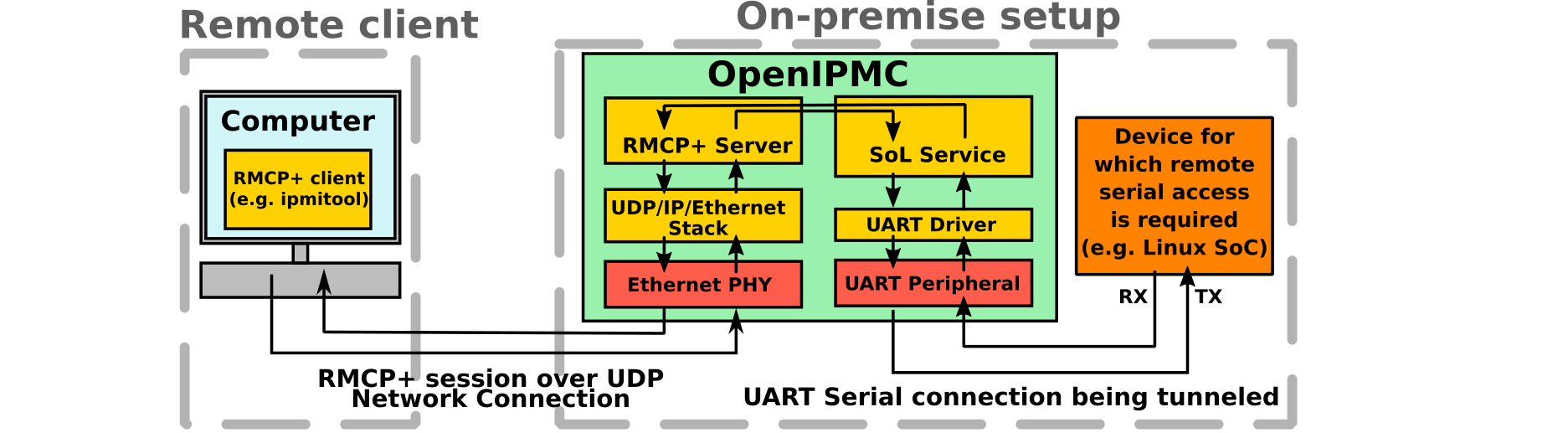}
  \caption{\corrtex{Functional scheme of the Serial-over-LAN application case.}}\label{sol-uart-over-rmcpplus}
\end{figure}

ATCA boards targeted by the OpenIPMC project, such as Serenity and Apollo, feature an onboard embedded computer running Linux to perform high-level tasks such as detector calibration. To facilitate the management of this computer when it has problems booting or is no longer directly accessible from the network, it is desirable to have remote access to its serial terminal, exposed on one of its UART interfaces\,\cite{ZynqUSpTRM}. The IPMI v1.5 standard\,\cite{IPMIv1p5}, on which part of the ATCA standard is based, defines the use of the Remote Management Control Protocol (RMCP) to transport IPMI messages over the network, and the subsequent IPMI v2.0 standard\,\cite{IPMIv2p0} adds new IPMI features (RMCP+) to RMCP, including the possibility of tunneling the input and output of a serial line over a network session, a feature known as \emph{Serial-over-LAN (SoL)} and shown on Fig. \ref{sol-uart-over-rmcpplus}.
OpenIPMC-FW has been updated to support RMCP+ SoL. This feature was tested in a bench top setup to monitor the boot log of a Raspberry Pi single-board computer and to access its Linux terminal via a remote connection established using the open source IPMI software package \emph{ipmitool}\,\cite{ipmitool}. The tests showed the SoL terminal session to work and helped to tune the buffer sizes to avoid dropped characters. Further tests are planned for Serenity and Apollo once the new revisions of these boards become available, providing the needed UART connections to the IPMC.

\subsection{YAFFS storage, TFTP server and configuration database}
\begin{figure}[h]
  \centering
  \includegraphics[height=35mm]{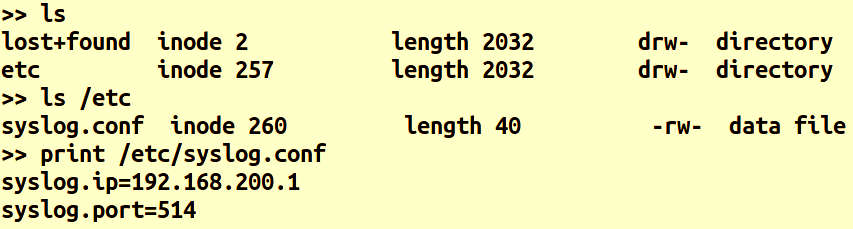}\\
  \caption{Browsing the content of the file system from the CLI.}
  \label{fig:yaffs_ls-4-rgb-large}
\end{figure}

A subset of the configuration settings for the IPMC may need to be frequently changed by the user (e.g. changing the Syslog server address or forcing a specific IP address for the device) and therefore need to be stored in RAM to allow its manipulation. At the same time, it is desirable that these settings are not wiped upon a system reset and, therefore, a mechanism should make them persistent. OpenIPMC-FW \corrtex{has been updated to support} the Yet Another Flash File System version 2 (YAFFS2)\,\cite{yaffs2} to mount a file system in a region of the external 128 MiB QSPI flash installed on the mezzanine PCB. The file system can be browsed from the IPMC Command Line Interface (CLI) and files can be moved, copied, deleted and edited (Fig.\,\ref{fig:yaffs_ls-4-rgb-large}). A \corrtex{Trivial File Transfer Protocol (TFTP)\,\cite{RFC1350}} server running in the firmware gives access to the mounted file system, such that files can be uploaded and downloaded remotely. Furthermore, OpenIPMC-FW saves run-time configuration settings in an in-memory dictionary, with entries which can be added, modified and deleted directly from the CLI by the user. The in-memory set of settings can also be saved into the file system as a set of human-readable files, which provides an excellent integration with the TFTP server. The settings stored in the file system are used to provide persistence and to bootstrap the in-memory configuration database after each reboot of the system.
\subsection{ClientID in DHCP}
\corrtex{In a large ATCA electronic system, such as the back-end of an HL-LHC experiment, it is useful to assign an unique address to each specific slot able to accommodate an electronic board (\emph{geographical addressing}), such that the identity and role of physically identical boards can be defined at runtime by the position they occupy. This organization helps the centralized management of the infrastructure and speeds up the replacement of failed boards, since the spares do not need to be pre-programmed with a specific identity and software before their insertion. The PICMG ATCA HPM.3\,\cite{picmg_hpm.3} standard defines a scheme to implement this geographical addressing using the} DHCP protocol\,\cite{RFC2131}, which \corrtex{is commonly used in networks to} auto-configure the connected devices\corrtex{. Following the HPM.3 specification, s}upport was added into OpenIPMC-FW for option 61 of the DHCP standard\,\cite{RFC2132} (ClientID), which allows \corrtex{a device to declare its identity using} an unique 20-byte \corrtex{name when using the DHCP protocol to request an IP address}. \corrtex{In the OpenIPMC-FW implementation, the device ClientID} is built from from the \corrtex{unique} ATCA shelf name and the slot position occupied by the board in the ATCA shelf, such that each slot corresponds to a well-defined IP address which can be used to remotely control and configure the board hosted into it.

\section{Conclusion and outlook}
In this article, we discussed the recent developments in the OpenIPMC project. Support for Xilinx Virtual Cable (XVC) to program FPGAs, Syslog for sending alert and debugging messages, RMCP/RMCP+ and Serial-Over-LAN (SoL) \corrtex{to remotely access an onboard UART interface}, and the creation of a configuration database are some of the most important firmware improvements. \corrtex{Most of these features have been implemented to fulfill specific requirements of the Serenity and Apollo ATCA boards, RMCP/RMCP+/SoL and the ClientID in DHCP being specified in the IPMI 2.0 standard from which PICMG 3.0 (ATCA) derives and the PICMG HPM.3 standard, respectively. Nevertheless, all the new features are commonly used in embedded computing applications and are expected to be available for use in applications beyond the aforementioned ATCA boards. For example, the XVC function in OpenIPMC enables remote programming and debugging for FPGA devices in experiments where physical access is restricted due to space or other constraints.}

\acknowledgments
The authors acknowledge the Fundação de Amparo à Pesquisa do Estado de São Paulo (FAPESP) for its financial support through grants number 18/18955-0 and 18/25225-9 and the Brazilian National Council for Scientific and Technological Development (CNPq) for its support through the INCT CERN-Brazil program \corrtex{(406672/2022-9)}. Marvin Fuchs acknowledges the support by the Doctoral School ``Karlsruhe School of Elementary and Astroparticle Physics: Science and Technology''. We thank the members of the CMS Phase-2 Tracker Upgrade Data Processing Systems group, and in particular G. Iles, M. Pesaresi, P. Wittich, and T. Williams for the help in defining the requirements for IPMCs used in the Phase-2 back-end boards.




\end{document}